\documentclass[prl,aps,10pt,floatfix,twocolumn,showpacs]{revtex4-1} 
\usepackage{color,soul}
\usepackage{graphicx}
\usepackage{amsmath, amsthm, amssymb,mathtools}
\usepackage{amsthm}
\usepackage{multirow}
\usepackage[normalem]{ulem}
\usepackage{soul}
\usepackage{newunicodechar}
\newunicodechar{ﬁ}{fi}
\newunicodechar{ﬀ}{ff}
\newcommand{\be}{\begin{equation}}
\newcommand{\ee}{\end{equation}}
\newcommand{\bea}{\begin{eqnarray}}
\newcommand{\eea}{\end{eqnarray}}
\usepackage{array, makecell} 
\usepackage{tablefootnote}

\begin{document}
\title{Bidisperse ring polymers: topological glass to stacking}
\author{Projesh Kumar Roy}
\email{projeshkr@imsc.res.in}
\affiliation{The Institute of Mathematical Sciences, C.I.T. Campus,
Taramani, Chennai 600113, India}
\affiliation{Homi Bhabha National Institute, Training School Complex, Anushakti Nagar, Mumbai -- 400094, Maharashtra, India}

\author{Pinaki Chaudhuri}
\email{pinakic@imsc.res.in}
\affiliation{The Institute of Mathematical Sciences, C.I.T. Campus,
Taramani, Chennai 600113, India}
\affiliation{Homi Bhabha National Institute, Training School Complex, Anushakti Nagar, Mumbai -- 400094, Maharashtra, India}

\author{Satyavani Vemparala}
\email{vani@imsc.res.in}
\affiliation{The Institute of Mathematical Sciences, C.I.T. Campus,
Taramani, Chennai 600113, India}
\affiliation{Homi Bhabha National Institute, Training School Complex, Anushakti Nagar, Mumbai -- 400094, Maharashtra, India}

\date{\today}

\begin{abstract}
Via large-scale molecular dynamics simulations, we observe the melting of a topological glass of stiff ring polymers by incorporating flexible ring polymers, along an isobaric path. As more flexible ring polymers are introduced, cluster glass-like structures emerge in the stiffer ring polymers with reduced orthogonal threading. This eventually evolves to a stacked columnar structure at an increased fraction of flexible ring polymers. Depletion interactions between the stiff and flexible rings drive the stacking, contingent on the disparity in flexibility in the ring polymer mixture. 

\end{abstract}
\maketitle

The emergence of self-organized ordered structures in soft materials is well-established now~\cite{yethiraj2003colloidal,zoldesi2005synthesis,li2016assembly,van2006colloids}. Two primary routes that lead to such ordering are effective hardcore ~\cite{cheng1999controlled,royall2013search,bommineni2020spontaneous} and depletion interactions~\cite{young2013directional,lekkerkerker2011depletion,karas2016using,park2022direct,miyazaki2022asakura}. For example, various studies on mixtures of non-deformable hard sphere colloidal systems have shown that such systems exhibit rich phases including crystalline and amorphous states as a function of the ratio of hard sphere sizes~\cite{asakura1954interaction,dijkstra1998phase,dickman1997entropic,dijkstra1999direct,kobayashi2021critical,wu2021size,vermolen2009fabrication}. However, the exploration of parameter space for mixtures of hardcore objects is limited. Introducing deformability could substantially enhance the availability of viable structures and lead to the emergence of interesting non-equilibrium or kinetically trapped phases.

Ring polymers with easily controllable backbone stiffness have emerged as excellent model systems for understanding the emergence of phase behavior in soft and deformable colloidal systems, with tuneable interactions~\cite{mukhopadhyay2011packings, makse2000packing, batista2010crystallization, boromand2018jamming}. Recent studies have demonstrated the emergence of glass-like states for dense ring polymer systems, which are inaccessible to their linear counterparts~\cite{Michieletto_Turner_PNAS_2016, Michieletto_Rosa_PhysRevLett_2017, michieletto2017ring}. Known as topological glasses, these
glassy states are deemed to occur via the interpenetration of rings (termed as \textit{threading}) due to the existence of closed-loop topological constraints \cite{Obukhov_Duke_PhysRevLett_1994, Obukov_Colby_Macromol_1994,Tsalikis_Vlassopoulos_MacroLett_2016,Dell_Schweizer_SoftMatter_2018}. Moreover, we recently~\cite{roy2022effect} showed that by carefully tuning the ring polymer stiffness, glassy states become easily accessible even at lower densities, without any artificial interventions  \cite{Michieletto_Turner_PNAS_2016, Michieletto_Rosa_PhysRevLett_2017, michieletto2017ring}.

In this work, we investigate how the mixing of rings with varying stiffness affects the stability of topological glassy states. Our motivation stems from recent studies that explore such mixing effects in other deformable soft objects, such as linear polymers \cite{milchev2020entropic, egorov2021phase, milchev2021blends} and mixtures of soft and hard colloids~\cite{merola2018asymmetric,parisi2021effect}. These studies have demonstrated the significance of entropic effects in determining the emergent phase behavior~\cite{frenkel2015order}. Recent findings~\cite{egorov2022linear,chubak2021multiscale} also suggest that ring polymers could be stronger depleting agents than linear polymers, indicating the potential for exotic self-assembly of deformable objects via depletion interactions. 

In this study, we use extensive molecular dynamics (MD) simulations to report the emergence of novel ordered aggregates, in the form of columnar stacked structures, in a binary mixture of ring polymers with different stiffness parameters, via a demixing process. We investigate the conformational landscape of the formation of such soft ordered aggregates of stiffer rings by exploring the binary mixture ratio of different ring polymer stiffness. Furthermore, we examine the dynamics of these aggregates to gain insights into the interplay between depletion effects of deformability and dynamical arrest in topologically constrained ring polymer systems.

In this study, we consider a system of 400 non-concatenated ring polymers, where each polymer has 100 monomers i.e. a total of 40,000 particles. Similar to our previous work~\cite{roy2022effect}, we have modeled various interactions between the monomers based on Kremer-Grest (KG) {bead-spring} model~\cite{Kurt_Grest_JChemPhys_1990, Halverson_Kremer_JChemPhys_2011, Halverson_Kremer_JChemPhys_2011_2}, which consists of--(a) WCA potential~\cite{Weeks_Andersen_JChemPhys_1971} for non-bonded interactions between any two monomers, (b) FENE potential~\cite{Kurt_Grest_JChemPhys_1990}, for bonded interactions, and (c) Kratky-Porod type angular potential~\cite{Doi_Edwards, Guo_Zhang_Polymers_2020}, which controls the stiffness of the ring polymer. Reduced LJ units are used through out the simulations. Details of MD simulations, via LAMMPS~\cite{LAMMPS}, can be found in the Supplementary Information (SI) \cite{si}, section S.I and S.II. 
\begin{figure*}
\centering
\includegraphics[scale=0.7]{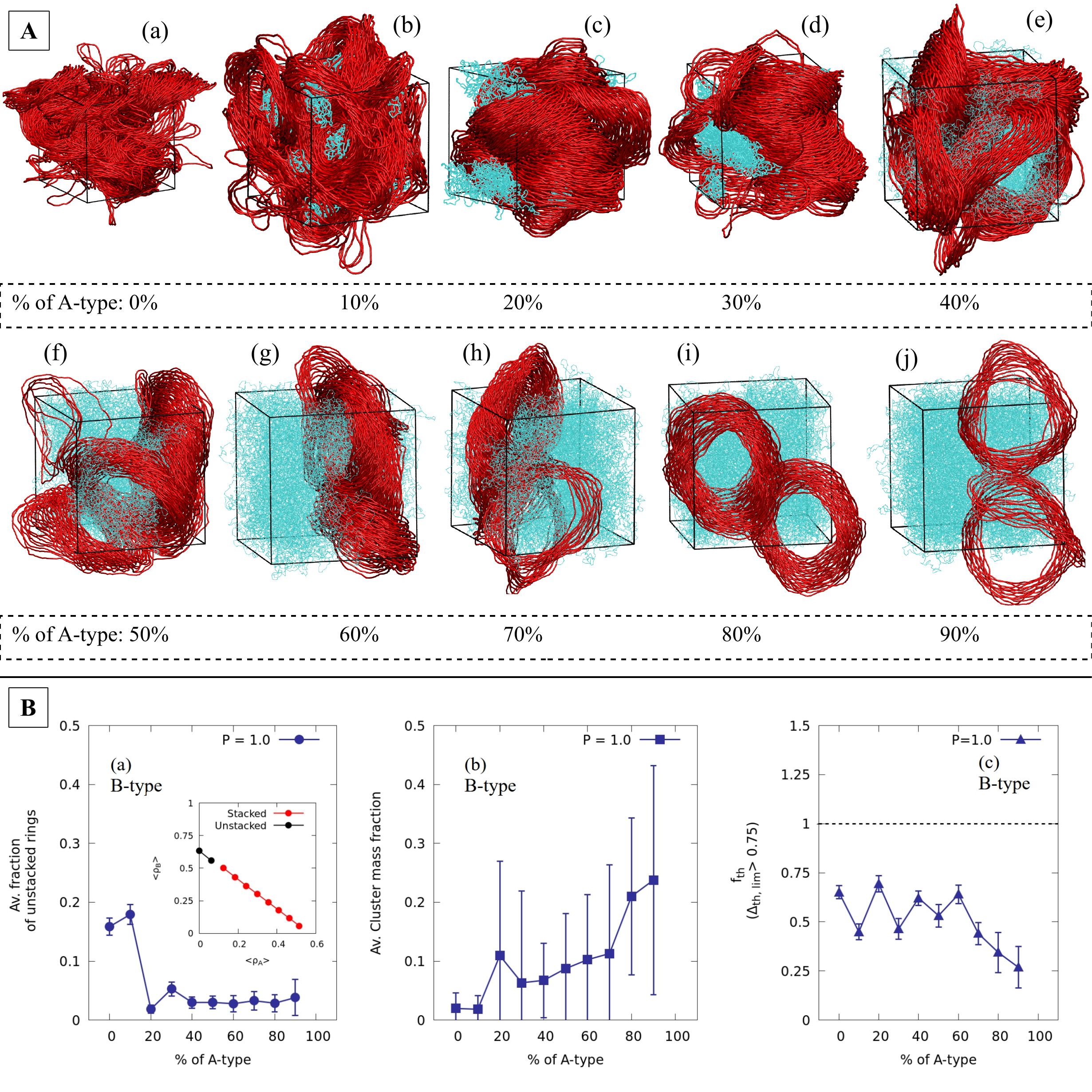}
\caption{(A)(a--j) Snapshots of the binary mixture of ring polymers having stiffness $K^A_{\theta}$=1.0  and $K^B_{\theta}$=20.0  at ambient pressure $P=1.0$  and temperature $T=1.0$, labeled via concentration of A-type rings. Color code: Cyan=A-type, Red=B-type rings. (B) With changing concentration of A-type rings, (a) variation of the average fraction of unstacked B-type rings. Inset: A phase diagram in the space of the obtained average densities of A-type and B-type rings, with the regime of stacking behavior marked. (b) the average mass fraction in clustered form (c) variation of the fraction of threaded rings (with $\Delta_{\text{th,lim}} > 0.75$) of B-type rings.}
\label{fig:snapshot}
\end{figure*}

The starting point of our study is a state point studied in our previous work~\cite{roy2022effect}, where a system of rings having a stiffness $K_{\theta}$ = 20.0 showed glassy dynamics at ambient pressure of $P=1.0$  and temperature $T=1.0$. Such a glassy state has a topological nature, i.e. the glassiness primarily arises from the threading of the rings. A snapshot of such a structure is shown in Fig.~\ref{fig:snapshot}.A(a), where the disorderedness of the assembly is evident (also see movie1 in SI \cite{si}). Our objective is to understand how the insertion of the more flexible rings changes the structural and dynamical properties of the system, with changing composition of the binary mixture at a fixed pressure and temperature. For the main part of our discussions, we focus on the mixture consisting of highly contrasting stiffness, {\it viz.} $K^A_{\theta}$=1.0  and $K^B_{\theta}$=20.0 . For these parameters, at $P=1.0$ , the stiffer single polymer takes the shape of a nearly planar ring, whereas the more flexible polymer has a compact globular structure. Further, for the A-type=100\% state, i.e. where only the highly flexible rings are present, the system is in a liquid state at this pressure~\cite{roy2022effect}. Thus, we are probing what are the intermediate states in the A-B phase diagram, as we vary the composition along an isobaric path. Later, we also discuss how the variation of $K^A_{\theta}$  in the mixture affects the observed behavior.

In Fig.~\ref{fig:snapshot}.A(a--j), we show the emergent structures, sampled at long times, via the interplay of the rings having two different stiffness parameters. The striking finding is that the disordered structure formed by the stiffer B-type rings in the absence of A-type flexible rings evolves into a completely different structure with increasing fractions of the more flexible rings. At a small concentration of the A-type rings (e.g. $10\%$), the emergence of stacking is visible, i.e. partial axial ordering of the stiffer rings is observed due to the intermixing of the two species (see movie2 in SI \cite{si}). With increasing concentration of the A-type rings, these stacks become more well-pronounced, and eventually at the dilute composition of the B-type rings, when \% of A-type rings $\gtrsim$ 70 \% (see Fig.~\ref{fig:snapshot}.A(h-j)), two stacks emerge which should eventually merge to form a single stack if kinetic trapping is overcome. Indeed, we have verified that at this state point a single stack is the stable equilibrium structure; see SI \cite{si}, section S.IV, Fig.S.5 and movie3 for further details. Overall, the gallery in Fig.~\ref{fig:snapshot} clearly demonstrates the melting of the topological glass formed by the stiffer rings via the increasing insertion of the more flexible rings. A subsequent de-mixing transition between the two species eventually leads to B-type rings forming more ordered structures in the form of columnar stacks. 

To quantify the visible stacking behavior of the B-type rings (see Fig.\ref{fig:snapshot}), we use the algorithm by Poier \textit{et al}~\cite{Poier_Blaak_SoftMatter_2016, Poier_Blaak_Macromol_2015} which allows us to identify neighboring rings that stack upon each other and thereby form a cluster; see SI \cite{si}, section S.I.C for the details of this algorithm. Via this process, we can also quantify the number of the unstacked stiff rings, which shows a marked decrease beyond a certain composition (see Fig.~\ref{fig:snapshot}.B(a)), strongly suggesting that the stacking tendency increases with increasing concentration of A-type rings. Signatures of unstacking are also visible via a drop in the potential energy of the B-type rings, see SI \cite{si} Fig.S.2. This is summarized in the form of a phase diagram in the space of average densities of A-type and B-type rings (inset of Fig.~\ref{fig:snapshot}.B(a)), with the line marking the isobaric path that we traverse in our study. We further measure the extent of clustering among such stacks in the following way: a cluster, having a minimum size of two rings, is recognized when every member therein is well stacked with its neighbors (see SI \cite{si}, section S.I.C for details). In Fig.~\ref{fig:snapshot}.B(b), the average cluster size, normalized by the total number of B-type rings at that composition, is plotted, which shows that the average mass fraction of the B-type rings in a cluster progressively increases with increasing dilution, indicating that at large dilution, B-type rings form a single cluster (see SI \cite{si}, section S.IV). 

An important question is to understand how the introduction of flexible rings affects the threading among the stiffer rings, which is the basis of formation of the topological glass in the pure B-type system at $P=1.0$. We investigate this by measuring the average threading parameter ($\langle\Delta_{\text{th}}\rangle$) (see Ref.~\citenum{{roy2022effect}} for definition) among the stiff rings, and find a decrease in threading at high dilutions (see Fig.~\ref{fig:snapshot}.B(c)), with commensurate increase in stacking clusters.  Note that our threading detection algorithm neglects the peripheral threading of two flexible rings lying on top of each other, via coarse-graining (see SI \cite{si}, section S.V for details).

\begin{figure}[h]
\includegraphics[width=\columnwidth]{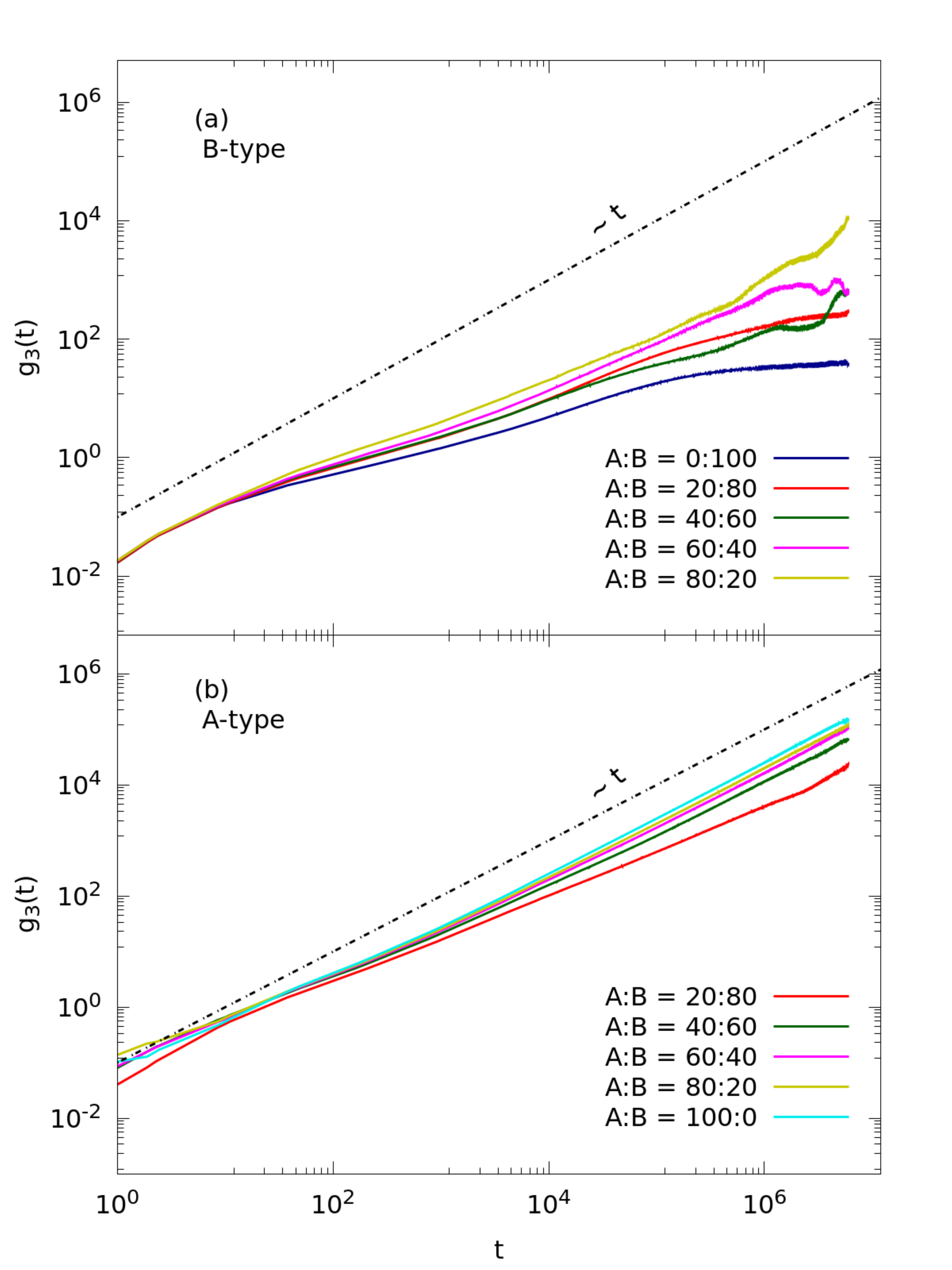}
\caption{Mean squared displacement of the center of mass, $g_3(t)$, at pressure $P=1.0$  for (a) B-type rings having $K^B_{\theta}$=20.0 , and (b) A-type rings having $K^A_{\theta}$=1.0 , at different compositions as labeled. In both panels, the dashed line corresponds to diffusive dynamics $\sim{t}$.}
\label{fig:msd_conc}
\end{figure}

So far, we have discussed the structural aspects of our observations with changing composition of binary mixtures. An important issue to address is whether the observed structures are in equilibrium or not. More specifically, what happens to the dynamically arrested state of the 100:0 state of the stiff B-type rings with increased insertion of the flexible A-type rings? These dynamical aspects are examined via the mean-square displacement (MSD), $g_3(t)$, of the center-of-mass of the rings. In Fig.~\ref{fig:msd_conc}, we show the MSD data for the two types of rings separately in sub-panels (a) and (b). If we focus on the stiffer B-type rings (see Fig.~\ref{fig:msd_conc}(a)), they display long-time dynamical arrest for the 100:0 case, as we have reported earlier~\cite{roy2022effect}. With the increased addition of the more flexible A-type rings, the dynamics still show long-time localization, i.e., the existence of a plateau where the plateau height increases with dilution. This is the case even for a 60:40 mixture. Thus, one can infer that the stiffer rings are still in a glassy state, i.e. the motion of the rings is still constrained although they gain some more space for fluctuating around. However, for the 80:20 case, we see the tendency towards long-time diffusion. Note that, even in the latter case, the long-time diffusion probably comes from the motion of the stacked cluster; but within the cluster, there are not much of structural rearrangements (see see movie3 in SI \cite{si}). For the flexible A-type rings (see Fig.~\ref{fig:msd_conc}(b)), we observe long-time diffusion in all cases. Thus, we infer that the insertion of these flexible rings leads to the initial coexistence of a glassy structure formed by the stiffer B-type rings with a liquid formed by the highly flexible A-type rings. These glassy structures are reminiscent of cluster glasses~\cite{slimani2014cluster,bernabei2013fluids,poier2015anisotropic,poier2016anisotropic}. Eventually, at higher concentrations of the A-type rings, the co-existing structure of the stiffer B-type rings emerges as an ordered structure of the stiffer rings.

We also note that when the fraction of A-type rings is small (e.g.  up to A:B $\sim$ 40:60), there is a pronounced transient sub-diffusive regime in their corresponding MSD, i.e.  $g_3(t)$ grows sub-linearly with time, before the eventual long-time diffusion, (see Fig.~\ref{fig:msd_conc}(b)). Such a transient behaviour is reminiscent of the motion of  interacting particles within an environment of quenched random soft obstacles \cite{schnyder2015rounding, schnyder2018crowding, vaibhav2022finite}. A similar constrained motion occurs here as well in this composition regime:  with the onset of stacking of the stiff B-type rings, albeit randomly oriented (see Fig.~\ref{fig:snapshot}), there emerges tubular channels creating a labyrinthine porous structure, which is the only accessible space for the flexible rings to move around  \footnote{The effective diameter of these tubes ($d_{\text{cav}}$) is proportional to the radius of gyration of the B-type rings ($R^B_g$), i.e., $d_{\text{cav}} \sim 2\times 14.5\sigma$, which is sufficiently large to allow the directed motion of the flexible rings}. With further dilution in the fraction of B-type rings leading to stacked ordering, more and more free space becomes available for the A-type rings, and therefore we only observe diffusive motion.

\begin{figure}[t]
\includegraphics[width=\columnwidth]{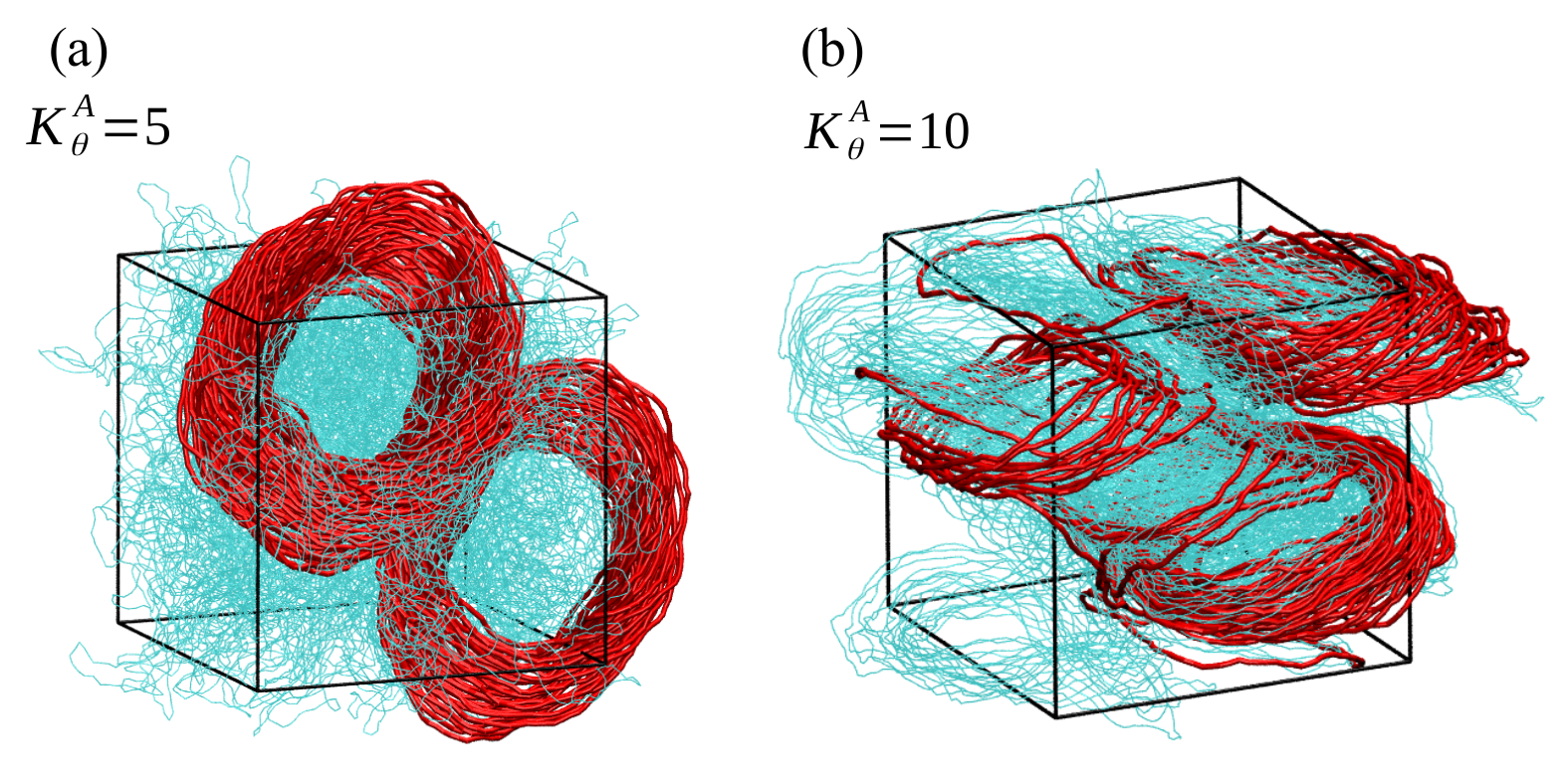}
\caption{Snapshots of the binary mixture with different $K^A_{\theta}$ values for A-type rings, {\it viz.} for $K^A_{\theta}$=5.0   (a) and $K^A_{\theta}$=20.0  for the 80:20 mixture at pressure $P=1.0$  and $K^B_{\theta}$=20.0 , with A-type rings shown in cyan and B-type rings shown in red.}
\label{fig:snapshot_K_variation}
\end{figure}

The demixing of the two species of rings within the binary mixture, seen here, is similar to the depletion-induced phase separations observed in other soft matter systems~\cite{tuinier2003depletion}. However, the induced ordered conformations of the stiffer ring polymers into clusters seen here are novel and suggest that depletion-induced phases can strongly depend on the asymmetry between the deformability of the constituent particles. For further insight into the role of asymmetry, we simulated systems with different stiffness ratios by tuning the flexibility of the A-type rings, e.g., $K^A_{\theta}$ = 5.0   and 10.0 , while keeping $K^B_{\theta}$ fixed at 20.0 . We study the effect of this variation at a specific compositional state point, viz.  A:B=80:20, where the stiffer B-type rings have a clear stacked structure.

As shown in Fig.~\ref{fig:snapshot_K_variation}, there is not much change in the observed structure of the clusters of B-type rings when $K^A_{\theta}$ is changed to 5.0  from 1.0 . However, for $K^A_{\theta}$ = 10.0 , the stack-clusters of B-type rings break apart. Hence, the depletion interactions between B-type rings are clearly diminished in the presence of stiffer A-type rings. In fact, for $K^A_{\theta}$ = 10.0 , we observe a competition for stacking and clustering between the A-type and B-type rings (see movie4 in SI \cite{si}). Earlier~\cite{roy2022effect} we have shown that, at the given pressure, $K^A_{\theta}$ = 10.0  systems take an elongated rod type shape in the bulk phase. In Fig~\ref{fig:snapshot_K_variation}(c), we observe that such rod-type rings can also form ordered stacks. This is quantified via an increase in stacking of less stiff polymers ($K^A_{\theta}$ = 10.0) and can be seen in  SI \cite{si}, section S.III and Fig. S.3. Therefore, in terms of stiffness difference, the critical value for observing a clear emergence of a coexisting ordered structure for the stiffer ring polymers lies in between 5.0  $\leq K^A_{\theta} \leq $ 10.0   for $K^B_{\theta}$ = 20.0   for the present study. 

In summary, we have used extensive molecular dynamics simulations to investigate the phase behavior of a binary mixture of ring polymers with contrasting stiffness at a fixed pressure. Starting from a topological glass of stiff ring polymers at an ambient pressure of $P=1.0$, we have demonstrated that the entangled disordered state unravels along the isobaric path and the rings self-organize to form an ordered stack with the increasing addition of highly flexible ring polymers. In the process, the orthogonal threading, which drives the formation of the topological glass, diminishes and nearly vanishes upon dilution of the stiffer rings. At intermediate compositions, our dynamical analysis reveals that a possible cluster glass is formed by the stiffer rings, with stacks of smaller sizes oriented in different directions. In this regime, the dynamics of the more flexible rings show an interesting transient sub-diffusion as they venture through the labyrinth formed by the disordered assembly of stacks. In the limit of a vanishing fraction of the stiff rings, a liquid state is formed by the flexible rings, as reported earlier \cite{roy2022effect}.

The interesting observation is that, unlike typical demixing scenarios where there is a well-defined spatial segregation between the demixed states, such as even in the case of a linear polymer mixture with different stiffness~\cite{milchev2020entropic, egorov2021phase, milchev2021blends}, the more flexible rings in this case populate the interior of the columnar phase formed by the stiff rings. The energetics of the stabilization of the columnar stack and the consequent interface with the liquid formed by the flexible rings need to be studied in future research. Additionally, we demonstrate that the ordered stacking in the form of a singular cluster becomes unstable when the asymmetry in the stiffness of the rings is decreased. Instead, we seem to obtain islands of such stacks, suggesting possible micro-demixing.

Our work is the first exploration of the phase space for a binary mixture of rings having different stiffness, and investigating how the variation in stiffness ratio would alter the observed structures. Via the trajectory along an isobaric path, we were able to identify the composition and pressure at which the coexistence of ordered stacking occurs with that of the liquid. Further studies are needed (e.g. along isochoric paths) to not only obtain the stability bounds of these structures in the phase diagram but also to chart out the overall phase behavior in the entire parameter space. For example, it remains to be investigated what other demixing scenarios are possible at different flexibility and sizes of the rings. Also, at larger pressures, the flexible rings too undergo a dynamical arrest and it would be interesting to explore its interplay with the glassy states formed by the topological glass, albeit these regimes would be highly non-equilibrium in nature. 
\begin{acknowledgements}
We thank the HPC facility at the Institute of Mathematical Sciences for computing time. PC and SV also acknowledge support via the sub-project on the Modeling of Soft Materials within the IMSC Apex project, funded by Department of Atomic Energy. We also thank J. Horbach for the interesting discussions.
\end{acknowledgements}


\begin{thebibliography}{58}%
\makeatletter
\providecommand \@ifxundefined [1]{%
 \@ifx{#1\undefined}
}%
\providecommand \@ifnum [1]{%
 \ifnum #1\expandafter \@firstoftwo
 \else \expandafter \@secondoftwo
 \fi
}%
\providecommand \@ifx [1]{%
 \ifx #1\expandafter \@firstoftwo
 \else \expandafter \@secondoftwo
 \fi
}%
\providecommand \natexlab [1]{#1}%
\providecommand \enquote  [1]{``#1''}%
\providecommand \bibnamefont  [1]{#1}%
\providecommand \bibfnamefont [1]{#1}%
\providecommand \citenamefont [1]{#1}%
\providecommand \href@noop [0]{\@secondoftwo}%
\providecommand \href [0]{\begingroup \@sanitize@url \@href}%
\providecommand \@href[1]{\@@startlink{#1}\@@href}%
\providecommand \@@href[1]{\endgroup#1\@@endlink}%
\providecommand \@sanitize@url [0]{\catcode `\\12\catcode `\$12\catcode
  `\&12\catcode `\#12\catcode `\^12\catcode `\_12\catcode `\%12\relax}%
\providecommand \@@startlink[1]{}%
\providecommand \@@endlink[0]{}%
\providecommand \url  [0]{\begingroup\@sanitize@url \@url }%
\providecommand \@url [1]{\endgroup\@href {#1}{\urlprefix }}%
\providecommand \urlprefix  [0]{URL }%
\providecommand \Eprint [0]{\href }%
\providecommand \doibase [0]{http://dx.doi.org/}%
\providecommand \selectlanguage [0]{\@gobble}%
\providecommand \bibinfo  [0]{\@secondoftwo}%
\providecommand \bibfield  [0]{\@secondoftwo}%
\providecommand \translation [1]{[#1]}%
\providecommand \BibitemOpen [0]{}%
\providecommand \bibitemStop [0]{}%
\providecommand \bibitemNoStop [0]{.\EOS\space}%
\providecommand \EOS [0]{\spacefactor3000\relax}%
\providecommand \BibitemShut  [1]{\csname bibitem#1\endcsname}%
\let\auto@bib@innerbib\@empty
\bibitem [{\citenamefont {Yethiraj}\ and\ \citenamefont {van
  Blaaderen}(2003)}]{yethiraj2003colloidal}%
  \BibitemOpen
  \bibfield  {author} {\bibinfo {author} {\bibfnamefont {A.}~\bibnamefont
  {Yethiraj}}\ and\ \bibinfo {author} {\bibfnamefont {A.}~\bibnamefont {van
  Blaaderen}},\ }\href@noop {} {\bibfield  {journal} {\bibinfo  {journal}
  {nature}\ }\textbf {\bibinfo {volume} {421}},\ \bibinfo {pages} {513}
  (\bibinfo {year} {2003})}\BibitemShut {NoStop}%
\bibitem [{\citenamefont {Zoldesi}\ and\ \citenamefont
  {Imhof}(2005)}]{zoldesi2005synthesis}%
  \BibitemOpen
  \bibfield  {author} {\bibinfo {author} {\bibfnamefont {C.~I.}\ \bibnamefont
  {Zoldesi}}\ and\ \bibinfo {author} {\bibfnamefont {A.}~\bibnamefont
  {Imhof}},\ }\href@noop {} {\bibfield  {journal} {\bibinfo  {journal}
  {Advanced Materials}\ }\textbf {\bibinfo {volume} {17}},\ \bibinfo {pages}
  {924} (\bibinfo {year} {2005})}\BibitemShut {NoStop}%
\bibitem [{\citenamefont {Li}\ \emph {et~al.}(2016)\citenamefont {Li},
  \citenamefont {Zhou},\ and\ \citenamefont {Han}}]{li2016assembly}%
  \BibitemOpen
  \bibfield  {author} {\bibinfo {author} {\bibfnamefont {B.}~\bibnamefont
  {Li}}, \bibinfo {author} {\bibfnamefont {D.}~\bibnamefont {Zhou}}, \ and\
  \bibinfo {author} {\bibfnamefont {Y.}~\bibnamefont {Han}},\ }\href@noop {}
  {\bibfield  {journal} {\bibinfo  {journal} {Nature Reviews Materials}\
  }\textbf {\bibinfo {volume} {1}},\ \bibinfo {pages} {1} (\bibinfo {year}
  {2016})}\BibitemShut {NoStop}%
\bibitem [{\citenamefont {van Blaaderen}(2006)}]{van2006colloids}%
  \BibitemOpen
  \bibfield  {author} {\bibinfo {author} {\bibfnamefont {A.}~\bibnamefont {van
  Blaaderen}},\ }\href@noop {} {\bibfield  {journal} {\bibinfo  {journal}
  {Nature}\ }\textbf {\bibinfo {volume} {439}},\ \bibinfo {pages} {545}
  (\bibinfo {year} {2006})}\BibitemShut {NoStop}%
\bibitem [{\citenamefont {Cheng}\ \emph {et~al.}(1999)\citenamefont {Cheng},
  \citenamefont {Russel},\ and\ \citenamefont {Chaikin}}]{cheng1999controlled}%
  \BibitemOpen
  \bibfield  {author} {\bibinfo {author} {\bibfnamefont {Z.}~\bibnamefont
  {Cheng}}, \bibinfo {author} {\bibfnamefont {W.~B.}\ \bibnamefont {Russel}}, \
  and\ \bibinfo {author} {\bibfnamefont {P.}~\bibnamefont {Chaikin}},\
  }\href@noop {} {\bibfield  {journal} {\bibinfo  {journal} {Nature}\ }\textbf
  {\bibinfo {volume} {401}},\ \bibinfo {pages} {893} (\bibinfo {year}
  {1999})}\BibitemShut {NoStop}%
\bibitem [{\citenamefont {Royall}\ \emph {et~al.}(2013)\citenamefont {Royall},
  \citenamefont {Poon},\ and\ \citenamefont {Weeks}}]{royall2013search}%
  \BibitemOpen
  \bibfield  {author} {\bibinfo {author} {\bibfnamefont {C.~P.}\ \bibnamefont
  {Royall}}, \bibinfo {author} {\bibfnamefont {W.~C.}\ \bibnamefont {Poon}}, \
  and\ \bibinfo {author} {\bibfnamefont {E.~R.}\ \bibnamefont {Weeks}},\
  }\href@noop {} {\bibfield  {journal} {\bibinfo  {journal} {Soft Matter}\
  }\textbf {\bibinfo {volume} {9}},\ \bibinfo {pages} {17} (\bibinfo {year}
  {2013})}\BibitemShut {NoStop}%
\bibitem [{\citenamefont {Bommineni}\ \emph {et~al.}(2020)\citenamefont
  {Bommineni}, \citenamefont {Klement},\ and\ \citenamefont
  {Engel}}]{bommineni2020spontaneous}%
  \BibitemOpen
  \bibfield  {author} {\bibinfo {author} {\bibfnamefont {P.~K.}\ \bibnamefont
  {Bommineni}}, \bibinfo {author} {\bibfnamefont {M.}~\bibnamefont {Klement}},
  \ and\ \bibinfo {author} {\bibfnamefont {M.}~\bibnamefont {Engel}},\
  }\href@noop {} {\bibfield  {journal} {\bibinfo  {journal} {Physical Review
  Letters}\ }\textbf {\bibinfo {volume} {124}},\ \bibinfo {pages} {218003}
  (\bibinfo {year} {2020})}\BibitemShut {NoStop}%
\bibitem [{\citenamefont {Young}\ \emph {et~al.}(2013)\citenamefont {Young},
  \citenamefont {Personick}, \citenamefont {Engel}, \citenamefont {Damasceno},
  \citenamefont {Barnaby}, \citenamefont {Bleher}, \citenamefont {Li},
  \citenamefont {Glotzer}, \citenamefont {Lee},\ and\ \citenamefont
  {Mirkin}}]{young2013directional}%
  \BibitemOpen
  \bibfield  {author} {\bibinfo {author} {\bibfnamefont {K.~L.}\ \bibnamefont
  {Young}}, \bibinfo {author} {\bibfnamefont {M.~L.}\ \bibnamefont
  {Personick}}, \bibinfo {author} {\bibfnamefont {M.}~\bibnamefont {Engel}},
  \bibinfo {author} {\bibfnamefont {P.~F.}\ \bibnamefont {Damasceno}}, \bibinfo
  {author} {\bibfnamefont {S.~N.}\ \bibnamefont {Barnaby}}, \bibinfo {author}
  {\bibfnamefont {R.}~\bibnamefont {Bleher}}, \bibinfo {author} {\bibfnamefont
  {T.}~\bibnamefont {Li}}, \bibinfo {author} {\bibfnamefont {S.~C.}\
  \bibnamefont {Glotzer}}, \bibinfo {author} {\bibfnamefont {B.}~\bibnamefont
  {Lee}}, \ and\ \bibinfo {author} {\bibfnamefont {C.~A.}\ \bibnamefont
  {Mirkin}},\ }\href@noop {} {\bibfield  {journal} {\bibinfo  {journal}
  {Angewandte Chemie International Edition}\ }\textbf {\bibinfo {volume}
  {52}},\ \bibinfo {pages} {13980} (\bibinfo {year} {2013})}\BibitemShut
  {NoStop}%
\bibitem [{\citenamefont {Lekkerkerker}\ \emph {et~al.}(2011)\citenamefont
  {Lekkerkerker}, \citenamefont {Tuinier}, \citenamefont {Lekkerkerker},\ and\
  \citenamefont {Tuinier}}]{lekkerkerker2011depletion}%
  \BibitemOpen
  \bibfield  {author} {\bibinfo {author} {\bibfnamefont {H.~N.}\ \bibnamefont
  {Lekkerkerker}}, \bibinfo {author} {\bibfnamefont {R.}~\bibnamefont
  {Tuinier}}, \bibinfo {author} {\bibfnamefont {H.~N.}\ \bibnamefont
  {Lekkerkerker}}, \ and\ \bibinfo {author} {\bibfnamefont {R.}~\bibnamefont
  {Tuinier}},\ }\href@noop {} {\emph {\bibinfo {title} {Depletion
  interaction}}}\ (\bibinfo  {publisher} {Springer},\ \bibinfo {year}
  {2011})\BibitemShut {NoStop}%
\bibitem [{\citenamefont {Karas}\ \emph {et~al.}(2016)\citenamefont {Karas},
  \citenamefont {Glaser},\ and\ \citenamefont {Glotzer}}]{karas2016using}%
  \BibitemOpen
  \bibfield  {author} {\bibinfo {author} {\bibfnamefont {A.~S.}\ \bibnamefont
  {Karas}}, \bibinfo {author} {\bibfnamefont {J.}~\bibnamefont {Glaser}}, \
  and\ \bibinfo {author} {\bibfnamefont {S.~C.}\ \bibnamefont {Glotzer}},\
  }\href@noop {} {\bibfield  {journal} {\bibinfo  {journal} {Soft Matter}\
  }\textbf {\bibinfo {volume} {12}},\ \bibinfo {pages} {5199} (\bibinfo {year}
  {2016})}\BibitemShut {NoStop}%
\bibitem [{\citenamefont {Park}\ \emph {et~al.}(2022)\citenamefont {Park},
  \citenamefont {Hwang},\ and\ \citenamefont {Kim}}]{park2022direct}%
  \BibitemOpen
  \bibfield  {author} {\bibinfo {author} {\bibfnamefont {S.}~\bibnamefont
  {Park}}, \bibinfo {author} {\bibfnamefont {H.}~\bibnamefont {Hwang}}, \ and\
  \bibinfo {author} {\bibfnamefont {S.-H.}\ \bibnamefont {Kim}},\ }\href@noop
  {} {\bibfield  {journal} {\bibinfo  {journal} {Journal of the American
  Chemical Society}\ }\textbf {\bibinfo {volume} {144}},\ \bibinfo {pages}
  {18397} (\bibinfo {year} {2022})}\BibitemShut {NoStop}%
\bibitem [{\citenamefont {Miyazaki}\ \emph {et~al.}(2022)\citenamefont
  {Miyazaki}, \citenamefont {Schweizer}, \citenamefont {Thirumalai},
  \citenamefont {Tuinier},\ and\ \citenamefont
  {Zaccarelli}}]{miyazaki2022asakura}%
  \BibitemOpen
  \bibfield  {author} {\bibinfo {author} {\bibfnamefont {K.}~\bibnamefont
  {Miyazaki}}, \bibinfo {author} {\bibfnamefont {K.}~\bibnamefont {Schweizer}},
  \bibinfo {author} {\bibfnamefont {D.}~\bibnamefont {Thirumalai}}, \bibinfo
  {author} {\bibfnamefont {R.}~\bibnamefont {Tuinier}}, \ and\ \bibinfo
  {author} {\bibfnamefont {E.}~\bibnamefont {Zaccarelli}},\ }\href@noop {}
  {\enquote {\bibinfo {title} {The asakura--oosawa theory: Entropic forces in
  physics, biology, and soft matter},}\ } (\bibinfo {year} {2022})\BibitemShut
  {NoStop}%
\bibitem [{\citenamefont {Asakura}\ and\ \citenamefont
  {Oosawa}(1954)}]{asakura1954interaction}%
  \BibitemOpen
  \bibfield  {author} {\bibinfo {author} {\bibfnamefont {S.}~\bibnamefont
  {Asakura}}\ and\ \bibinfo {author} {\bibfnamefont {F.}~\bibnamefont
  {Oosawa}},\ }\href@noop {} {\bibfield  {journal} {\bibinfo  {journal} {The
  Journal of chemical physics}\ }\textbf {\bibinfo {volume} {22}},\ \bibinfo
  {pages} {1255} (\bibinfo {year} {1954})}\BibitemShut {NoStop}%
\bibitem [{\citenamefont {Dijkstra}\ \emph {et~al.}(1998)\citenamefont
  {Dijkstra}, \citenamefont {van Roij},\ and\ \citenamefont
  {Evans}}]{dijkstra1998phase}%
  \BibitemOpen
  \bibfield  {author} {\bibinfo {author} {\bibfnamefont {M.}~\bibnamefont
  {Dijkstra}}, \bibinfo {author} {\bibfnamefont {R.}~\bibnamefont {van Roij}},
  \ and\ \bibinfo {author} {\bibfnamefont {R.}~\bibnamefont {Evans}},\
  }\href@noop {} {\bibfield  {journal} {\bibinfo  {journal} {Physical review
  letters}\ }\textbf {\bibinfo {volume} {81}},\ \bibinfo {pages} {2268}
  (\bibinfo {year} {1998})}\BibitemShut {NoStop}%
\bibitem [{\citenamefont {Dickman}\ \emph {et~al.}(1997)\citenamefont
  {Dickman}, \citenamefont {Attard},\ and\ \citenamefont
  {Simonian}}]{dickman1997entropic}%
  \BibitemOpen
  \bibfield  {author} {\bibinfo {author} {\bibfnamefont {R.}~\bibnamefont
  {Dickman}}, \bibinfo {author} {\bibfnamefont {P.}~\bibnamefont {Attard}}, \
  and\ \bibinfo {author} {\bibfnamefont {V.}~\bibnamefont {Simonian}},\
  }\href@noop {} {\bibfield  {journal} {\bibinfo  {journal} {The Journal of
  chemical physics}\ }\textbf {\bibinfo {volume} {107}},\ \bibinfo {pages}
  {205} (\bibinfo {year} {1997})}\BibitemShut {NoStop}%
\bibitem [{\citenamefont {Dijkstra}\ \emph {et~al.}(1999)\citenamefont
  {Dijkstra}, \citenamefont {van Roij},\ and\ \citenamefont
  {Evans}}]{dijkstra1999direct}%
  \BibitemOpen
  \bibfield  {author} {\bibinfo {author} {\bibfnamefont {M.}~\bibnamefont
  {Dijkstra}}, \bibinfo {author} {\bibfnamefont {R.}~\bibnamefont {van Roij}},
  \ and\ \bibinfo {author} {\bibfnamefont {R.}~\bibnamefont {Evans}},\
  }\href@noop {} {\bibfield  {journal} {\bibinfo  {journal} {Physical review
  letters}\ }\textbf {\bibinfo {volume} {82}},\ \bibinfo {pages} {117}
  (\bibinfo {year} {1999})}\BibitemShut {NoStop}%
\bibitem [{\citenamefont {Kobayashi}\ \emph {et~al.}(2021)\citenamefont
  {Kobayashi}, \citenamefont {Rohrbach}, \citenamefont {Scheichl},
  \citenamefont {Wilding},\ and\ \citenamefont {Jack}}]{kobayashi2021critical}%
  \BibitemOpen
  \bibfield  {author} {\bibinfo {author} {\bibfnamefont {H.}~\bibnamefont
  {Kobayashi}}, \bibinfo {author} {\bibfnamefont {P.~B.}\ \bibnamefont
  {Rohrbach}}, \bibinfo {author} {\bibfnamefont {R.}~\bibnamefont {Scheichl}},
  \bibinfo {author} {\bibfnamefont {N.~B.}\ \bibnamefont {Wilding}}, \ and\
  \bibinfo {author} {\bibfnamefont {R.~L.}\ \bibnamefont {Jack}},\ }\href@noop
  {} {\bibfield  {journal} {\bibinfo  {journal} {Physical Review E}\ }\textbf
  {\bibinfo {volume} {104}},\ \bibinfo {pages} {044603} (\bibinfo {year}
  {2021})}\BibitemShut {NoStop}%
\bibitem [{\citenamefont {Wu}\ \emph {et~al.}(2021)\citenamefont {Wu},
  \citenamefont {Song}, \citenamefont {Chen}, \citenamefont {Song},
  \citenamefont {Porcar},\ and\ \citenamefont {Wang}}]{wu2021size}%
  \BibitemOpen
  \bibfield  {author} {\bibinfo {author} {\bibfnamefont {H.}~\bibnamefont
  {Wu}}, \bibinfo {author} {\bibfnamefont {K.}~\bibnamefont {Song}}, \bibinfo
  {author} {\bibfnamefont {W.-R.}\ \bibnamefont {Chen}}, \bibinfo {author}
  {\bibfnamefont {J.}~\bibnamefont {Song}}, \bibinfo {author} {\bibfnamefont
  {L.}~\bibnamefont {Porcar}}, \ and\ \bibinfo {author} {\bibfnamefont
  {Z.}~\bibnamefont {Wang}},\ }\href@noop {} {\bibfield  {journal} {\bibinfo
  {journal} {Physical Review Research}\ }\textbf {\bibinfo {volume} {3}},\
  \bibinfo {pages} {033271} (\bibinfo {year} {2021})}\BibitemShut {NoStop}%
\bibitem [{\citenamefont {Vermolen}\ \emph {et~al.}(2009)\citenamefont
  {Vermolen}, \citenamefont {Kuijk}, \citenamefont {Filion}, \citenamefont
  {Hermes}, \citenamefont {Thijssen}, \citenamefont {Dijkstra},\ and\
  \citenamefont {Van~Blaaderen}}]{vermolen2009fabrication}%
  \BibitemOpen
  \bibfield  {author} {\bibinfo {author} {\bibfnamefont {E.}~\bibnamefont
  {Vermolen}}, \bibinfo {author} {\bibfnamefont {A.}~\bibnamefont {Kuijk}},
  \bibinfo {author} {\bibfnamefont {L.}~\bibnamefont {Filion}}, \bibinfo
  {author} {\bibfnamefont {M.}~\bibnamefont {Hermes}}, \bibinfo {author}
  {\bibfnamefont {J.}~\bibnamefont {Thijssen}}, \bibinfo {author}
  {\bibfnamefont {M.}~\bibnamefont {Dijkstra}}, \ and\ \bibinfo {author}
  {\bibfnamefont {A.}~\bibnamefont {Van~Blaaderen}},\ }\href@noop {} {\bibfield
   {journal} {\bibinfo  {journal} {Proceedings of the National Academy of
  Sciences}\ }\textbf {\bibinfo {volume} {106}},\ \bibinfo {pages} {16063}
  (\bibinfo {year} {2009})}\BibitemShut {NoStop}%
\bibitem [{\citenamefont {Mukhopadhyay}\ and\ \citenamefont
  {Peixinho}(2011)}]{mukhopadhyay2011packings}%
  \BibitemOpen
  \bibfield  {author} {\bibinfo {author} {\bibfnamefont {S.}~\bibnamefont
  {Mukhopadhyay}}\ and\ \bibinfo {author} {\bibfnamefont {J.}~\bibnamefont
  {Peixinho}},\ }\href@noop {} {\bibfield  {journal} {\bibinfo  {journal}
  {Physical Review E}\ }\textbf {\bibinfo {volume} {84}},\ \bibinfo {pages}
  {011302} (\bibinfo {year} {2011})}\BibitemShut {NoStop}%
\bibitem [{\citenamefont {Makse}\ \emph {et~al.}(2000)\citenamefont {Makse},
  \citenamefont {Johnson},\ and\ \citenamefont {Schwartz}}]{makse2000packing}%
  \BibitemOpen
  \bibfield  {author} {\bibinfo {author} {\bibfnamefont {H.~A.}\ \bibnamefont
  {Makse}}, \bibinfo {author} {\bibfnamefont {D.~L.}\ \bibnamefont {Johnson}},
  \ and\ \bibinfo {author} {\bibfnamefont {L.~M.}\ \bibnamefont {Schwartz}},\
  }\href@noop {} {\bibfield  {journal} {\bibinfo  {journal} {Physical review
  letters}\ }\textbf {\bibinfo {volume} {84}},\ \bibinfo {pages} {4160}
  (\bibinfo {year} {2000})}\BibitemShut {NoStop}%
\bibitem [{\citenamefont {Batista}\ and\ \citenamefont
  {Miller}(2010)}]{batista2010crystallization}%
  \BibitemOpen
  \bibfield  {author} {\bibinfo {author} {\bibfnamefont {V.~M.}\ \bibnamefont
  {Batista}}\ and\ \bibinfo {author} {\bibfnamefont {M.~A.}\ \bibnamefont
  {Miller}},\ }\href@noop {} {\bibfield  {journal} {\bibinfo  {journal}
  {Physical review letters}\ }\textbf {\bibinfo {volume} {105}},\ \bibinfo
  {pages} {088305} (\bibinfo {year} {2010})}\BibitemShut {NoStop}%
\bibitem [{\citenamefont {Boromand}\ \emph {et~al.}(2018)\citenamefont
  {Boromand}, \citenamefont {Signoriello}, \citenamefont {Ye}, \citenamefont
  {O?Hern},\ and\ \citenamefont {Shattuck}}]{boromand2018jamming}%
  \BibitemOpen
  \bibfield  {author} {\bibinfo {author} {\bibfnamefont {A.}~\bibnamefont
  {Boromand}}, \bibinfo {author} {\bibfnamefont {A.}~\bibnamefont
  {Signoriello}}, \bibinfo {author} {\bibfnamefont {F.}~\bibnamefont {Ye}},
  \bibinfo {author} {\bibfnamefont {C.~S.}\ \bibnamefont {O?Hern}}, \ and\
  \bibinfo {author} {\bibfnamefont {M.~D.}\ \bibnamefont {Shattuck}},\
  }\href@noop {} {\bibfield  {journal} {\bibinfo  {journal} {Physical review
  letters}\ }\textbf {\bibinfo {volume} {121}},\ \bibinfo {pages} {248003}
  (\bibinfo {year} {2018})}\BibitemShut {NoStop}%
\bibitem [{\citenamefont {Michieletto}\ and\ \citenamefont
  {Turner}(2016)}]{Michieletto_Turner_PNAS_2016}%
  \BibitemOpen
  \bibfield  {author} {\bibinfo {author} {\bibfnamefont {D.}~\bibnamefont
  {Michieletto}}\ and\ \bibinfo {author} {\bibfnamefont {M.~S.}\ \bibnamefont
  {Turner}},\ }\href@noop {} {\bibfield  {journal} {\bibinfo  {journal}
  {Proceedings of the National Academy of Sciences}\ }\textbf {\bibinfo
  {volume} {113}},\ \bibinfo {pages} {5195} (\bibinfo {year}
  {2016})}\BibitemShut {NoStop}%
\bibitem [{\citenamefont {Michieletto}\ \emph
  {et~al.}(2017{\natexlab{a}})\citenamefont {Michieletto}, \citenamefont
  {Nahali},\ and\ \citenamefont {Rosa}}]{Michieletto_Rosa_PhysRevLett_2017}%
  \BibitemOpen
  \bibfield  {author} {\bibinfo {author} {\bibfnamefont {D.}~\bibnamefont
  {Michieletto}}, \bibinfo {author} {\bibfnamefont {N.}~\bibnamefont {Nahali}},
  \ and\ \bibinfo {author} {\bibfnamefont {A.}~\bibnamefont {Rosa}},\
  }\href@noop {} {\bibfield  {journal} {\bibinfo  {journal} {Physical Review
  Letters}\ }\textbf {\bibinfo {volume} {119}},\ \bibinfo {pages} {197801}
  (\bibinfo {year} {2017}{\natexlab{a}})}\BibitemShut {NoStop}%
\bibitem [{\citenamefont {Michieletto}\ \emph
  {et~al.}(2017{\natexlab{b}})\citenamefont {Michieletto}, \citenamefont
  {Marenduzzo}, \citenamefont {Orlandini},\ and\ \citenamefont
  {Turner}}]{michieletto2017ring}%
  \BibitemOpen
  \bibfield  {author} {\bibinfo {author} {\bibfnamefont {D.}~\bibnamefont
  {Michieletto}}, \bibinfo {author} {\bibfnamefont {D.}~\bibnamefont
  {Marenduzzo}}, \bibinfo {author} {\bibfnamefont {E.}~\bibnamefont
  {Orlandini}}, \ and\ \bibinfo {author} {\bibfnamefont {M.~S.}\ \bibnamefont
  {Turner}},\ }\href@noop {} {\bibfield  {journal} {\bibinfo  {journal}
  {Polymers}\ }\textbf {\bibinfo {volume} {9}},\ \bibinfo {pages} {349}
  (\bibinfo {year} {2017}{\natexlab{b}})}\BibitemShut {NoStop}%
\bibitem [{\citenamefont {Obukhov}\ \emph
  {et~al.}(1994{\natexlab{a}})\citenamefont {Obukhov}, \citenamefont
  {Rubinstein},\ and\ \citenamefont {Duke}}]{Obukhov_Duke_PhysRevLett_1994}%
  \BibitemOpen
  \bibfield  {author} {\bibinfo {author} {\bibfnamefont {S.~P.}\ \bibnamefont
  {Obukhov}}, \bibinfo {author} {\bibfnamefont {M.}~\bibnamefont {Rubinstein}},
  \ and\ \bibinfo {author} {\bibfnamefont {T.}~\bibnamefont {Duke}},\ }\href
  {\doibase 10.1103/PhysRevLett.73.1263} {\bibfield  {journal} {\bibinfo
  {journal} {Phys. Rev. Lett.}\ }\textbf {\bibinfo {volume} {73}},\ \bibinfo
  {pages} {1263} (\bibinfo {year} {1994}{\natexlab{a}})}\BibitemShut {NoStop}%
\bibitem [{\citenamefont {Obukhov}\ \emph
  {et~al.}(1994{\natexlab{b}})\citenamefont {Obukhov}, \citenamefont
  {Rubinstein},\ and\ \citenamefont {Colby}}]{Obukov_Colby_Macromol_1994}%
  \BibitemOpen
  \bibfield  {author} {\bibinfo {author} {\bibfnamefont {S.~P.}\ \bibnamefont
  {Obukhov}}, \bibinfo {author} {\bibfnamefont {M.}~\bibnamefont {Rubinstein}},
  \ and\ \bibinfo {author} {\bibfnamefont {R.~H.}\ \bibnamefont {Colby}},\
  }\href {\doibase 10.1021/ma00090a012} {\bibfield  {journal} {\bibinfo
  {journal} {Macromolecules}\ }\textbf {\bibinfo {volume} {27}},\ \bibinfo
  {pages} {3191} (\bibinfo {year} {1994}{\natexlab{b}})}\BibitemShut {NoStop}%
\bibitem [{\citenamefont {Tsalikis}\ \emph {et~al.}(2016)\citenamefont
  {Tsalikis}, \citenamefont {Mavrantzas},\ and\ \citenamefont
  {Vlassopoulos}}]{Tsalikis_Vlassopoulos_MacroLett_2016}%
  \BibitemOpen
  \bibfield  {author} {\bibinfo {author} {\bibfnamefont {D.~G.}\ \bibnamefont
  {Tsalikis}}, \bibinfo {author} {\bibfnamefont {V.~G.}\ \bibnamefont
  {Mavrantzas}}, \ and\ \bibinfo {author} {\bibfnamefont {D.}~\bibnamefont
  {Vlassopoulos}},\ }\href {\doibase 10.1021/acsmacrolett.6b00259} {\bibfield
  {journal} {\bibinfo  {journal} {ACS Macro Letters}\ }\textbf {\bibinfo
  {volume} {5}},\ \bibinfo {pages} {755} (\bibinfo {year} {2016})},\ \Eprint
  {http://arxiv.org/abs/https://doi.org/10.1021/acsmacrolett.6b00259}
  {https://doi.org/10.1021/acsmacrolett.6b00259} \BibitemShut {NoStop}%
\bibitem [{\citenamefont {Dell}\ and\ \citenamefont
  {Schweizer}(2018)}]{Dell_Schweizer_SoftMatter_2018}%
  \BibitemOpen
  \bibfield  {author} {\bibinfo {author} {\bibfnamefont {Z.~E.}\ \bibnamefont
  {Dell}}\ and\ \bibinfo {author} {\bibfnamefont {K.~S.}\ \bibnamefont
  {Schweizer}},\ }\href {\doibase 10.1039/C8SM01722K} {\bibfield  {journal}
  {\bibinfo  {journal} {Soft Matter}\ }\textbf {\bibinfo {volume} {14}},\
  \bibinfo {pages} {9132} (\bibinfo {year} {2018})}\BibitemShut {NoStop}%
\bibitem [{\citenamefont {Roy}\ \emph {et~al.}(2022)\citenamefont {Roy},
  \citenamefont {Chaudhuri},\ and\ \citenamefont {Vemparala}}]{roy2022effect}%
  \BibitemOpen
  \bibfield  {author} {\bibinfo {author} {\bibfnamefont {P.~K.}\ \bibnamefont
  {Roy}}, \bibinfo {author} {\bibfnamefont {P.}~\bibnamefont {Chaudhuri}}, \
  and\ \bibinfo {author} {\bibfnamefont {S.}~\bibnamefont {Vemparala}},\
  }\href@noop {} {\bibfield  {journal} {\bibinfo  {journal} {Soft Matter}\
  }\textbf {\bibinfo {volume} {18}},\ \bibinfo {pages} {2959} (\bibinfo {year}
  {2022})}\BibitemShut {NoStop}%
\bibitem [{\citenamefont {Milchev}\ \emph {et~al.}(2020)\citenamefont
  {Milchev}, \citenamefont {Egorov}, \citenamefont {Midya}, \citenamefont
  {Binder},\ and\ \citenamefont {Nikoubashman}}]{milchev2020entropic}%
  \BibitemOpen
  \bibfield  {author} {\bibinfo {author} {\bibfnamefont {A.}~\bibnamefont
  {Milchev}}, \bibinfo {author} {\bibfnamefont {S.~A.}\ \bibnamefont {Egorov}},
  \bibinfo {author} {\bibfnamefont {J.}~\bibnamefont {Midya}}, \bibinfo
  {author} {\bibfnamefont {K.}~\bibnamefont {Binder}}, \ and\ \bibinfo {author}
  {\bibfnamefont {A.}~\bibnamefont {Nikoubashman}},\ }\href@noop {} {\bibfield
  {journal} {\bibinfo  {journal} {ACS Macro Letters}\ }\textbf {\bibinfo
  {volume} {9}},\ \bibinfo {pages} {1779} (\bibinfo {year} {2020})}\BibitemShut
  {NoStop}%
\bibitem [{\citenamefont {Egorov}\ \emph {et~al.}(2021)\citenamefont {Egorov},
  \citenamefont {Milchev}, \citenamefont {Nikoubashman},\ and\ \citenamefont
  {Binder}}]{egorov2021phase}%
  \BibitemOpen
  \bibfield  {author} {\bibinfo {author} {\bibfnamefont {S.~A.}\ \bibnamefont
  {Egorov}}, \bibinfo {author} {\bibfnamefont {A.}~\bibnamefont {Milchev}},
  \bibinfo {author} {\bibfnamefont {A.}~\bibnamefont {Nikoubashman}}, \ and\
  \bibinfo {author} {\bibfnamefont {K.}~\bibnamefont {Binder}},\ }\href@noop {}
  {\bibfield  {journal} {\bibinfo  {journal} {The Journal of Physical Chemistry
  B}\ } (\bibinfo {year} {2021})}\BibitemShut {NoStop}%
\bibitem [{\citenamefont {Milchev}\ \emph {et~al.}(2021)\citenamefont
  {Milchev}, \citenamefont {Egorov}, \citenamefont {Midya}, \citenamefont
  {Binder},\ and\ \citenamefont {Nikoubashman}}]{milchev2021blends}%
  \BibitemOpen
  \bibfield  {author} {\bibinfo {author} {\bibfnamefont {A.}~\bibnamefont
  {Milchev}}, \bibinfo {author} {\bibfnamefont {S.~A.}\ \bibnamefont {Egorov}},
  \bibinfo {author} {\bibfnamefont {J.}~\bibnamefont {Midya}}, \bibinfo
  {author} {\bibfnamefont {K.}~\bibnamefont {Binder}}, \ and\ \bibinfo {author}
  {\bibfnamefont {A.}~\bibnamefont {Nikoubashman}},\ }\href@noop {} {\bibfield
  {journal} {\bibinfo  {journal} {Polymers}\ }\textbf {\bibinfo {volume}
  {13}},\ \bibinfo {pages} {2270} (\bibinfo {year} {2021})}\BibitemShut
  {NoStop}%
\bibitem [{\citenamefont {Merola}\ \emph {et~al.}(2018)\citenamefont {Merola},
  \citenamefont {Parisi}, \citenamefont {Truzzolillo}, \citenamefont
  {Vlassopoulos}, \citenamefont {Deepak},\ and\ \citenamefont
  {Gauthier}}]{merola2018asymmetric}%
  \BibitemOpen
  \bibfield  {author} {\bibinfo {author} {\bibfnamefont {M.~C.}\ \bibnamefont
  {Merola}}, \bibinfo {author} {\bibfnamefont {D.}~\bibnamefont {Parisi}},
  \bibinfo {author} {\bibfnamefont {D.}~\bibnamefont {Truzzolillo}}, \bibinfo
  {author} {\bibfnamefont {D.}~\bibnamefont {Vlassopoulos}}, \bibinfo {author}
  {\bibfnamefont {V.~D.}\ \bibnamefont {Deepak}}, \ and\ \bibinfo {author}
  {\bibfnamefont {M.}~\bibnamefont {Gauthier}},\ }\href@noop {} {\bibfield
  {journal} {\bibinfo  {journal} {Journal of Rheology}\ }\textbf {\bibinfo
  {volume} {62}},\ \bibinfo {pages} {63} (\bibinfo {year} {2018})}\BibitemShut
  {NoStop}%
\bibitem [{\citenamefont {Parisi}\ \emph {et~al.}(2021)\citenamefont {Parisi},
  \citenamefont {Camargo}, \citenamefont {Makri}, \citenamefont {Gauthier},
  \citenamefont {Likos},\ and\ \citenamefont
  {Vlassopoulos}}]{parisi2021effect}%
  \BibitemOpen
  \bibfield  {author} {\bibinfo {author} {\bibfnamefont {D.}~\bibnamefont
  {Parisi}}, \bibinfo {author} {\bibfnamefont {M.}~\bibnamefont {Camargo}},
  \bibinfo {author} {\bibfnamefont {K.}~\bibnamefont {Makri}}, \bibinfo
  {author} {\bibfnamefont {M.}~\bibnamefont {Gauthier}}, \bibinfo {author}
  {\bibfnamefont {C.~N.}\ \bibnamefont {Likos}}, \ and\ \bibinfo {author}
  {\bibfnamefont {D.}~\bibnamefont {Vlassopoulos}},\ }\href@noop {} {\bibfield
  {journal} {\bibinfo  {journal} {The Journal of Chemical Physics}\ }\textbf
  {\bibinfo {volume} {155}},\ \bibinfo {pages} {034901} (\bibinfo {year}
  {2021})}\BibitemShut {NoStop}%
\bibitem [{\citenamefont {Frenkel}(2015)}]{frenkel2015order}%
  \BibitemOpen
  \bibfield  {author} {\bibinfo {author} {\bibfnamefont {D.}~\bibnamefont
  {Frenkel}},\ }\href@noop {} {\bibfield  {journal} {\bibinfo  {journal}
  {Nature materials}\ }\textbf {\bibinfo {volume} {14}},\ \bibinfo {pages} {9}
  (\bibinfo {year} {2015})}\BibitemShut {NoStop}%
\bibitem [{\citenamefont {Egorov}(2022)}]{egorov2022linear}%
  \BibitemOpen
  \bibfield  {author} {\bibinfo {author} {\bibfnamefont {S.~A.}\ \bibnamefont
  {Egorov}},\ }\href@noop {} {\bibfield  {journal} {\bibinfo  {journal}
  {Macromolecular Theory and Simulations}\ }\textbf {\bibinfo {volume} {31}},\
  \bibinfo {pages} {2100065} (\bibinfo {year} {2022})}\BibitemShut {NoStop}%
\bibitem [{\citenamefont {Chubak}\ \emph {et~al.}(2021)\citenamefont {Chubak},
  \citenamefont {Likos},\ and\ \citenamefont {Egorov}}]{chubak2021multiscale}%
  \BibitemOpen
  \bibfield  {author} {\bibinfo {author} {\bibfnamefont {I.}~\bibnamefont
  {Chubak}}, \bibinfo {author} {\bibfnamefont {C.~N.}\ \bibnamefont {Likos}}, \
  and\ \bibinfo {author} {\bibfnamefont {S.~A.}\ \bibnamefont {Egorov}},\
  }\href@noop {} {\bibfield  {journal} {\bibinfo  {journal} {The Journal of
  Physical Chemistry B}\ }\textbf {\bibinfo {volume} {125}},\ \bibinfo {pages}
  {4910} (\bibinfo {year} {2021})}\BibitemShut {NoStop}%
\bibitem [{\citenamefont {Kremer}\ and\ \citenamefont
  {Grest}(1990)}]{Kurt_Grest_JChemPhys_1990}%
  \BibitemOpen
  \bibfield  {author} {\bibinfo {author} {\bibfnamefont {K.}~\bibnamefont
  {Kremer}}\ and\ \bibinfo {author} {\bibfnamefont {G.~S.}\ \bibnamefont
  {Grest}},\ }\href {\doibase 10.1063/1.458541} {\bibfield  {journal} {\bibinfo
   {journal} {The Journal of Chemical Physics}\ }\textbf {\bibinfo {volume}
  {92}},\ \bibinfo {pages} {5057} (\bibinfo {year} {1990})},\ \Eprint
  {http://arxiv.org/abs/https://doi.org/10.1063/1.458541}
  {https://doi.org/10.1063/1.458541} \BibitemShut {NoStop}%
\bibitem [{\citenamefont {Halverson}\ \emph
  {et~al.}(2011{\natexlab{a}})\citenamefont {Halverson}, \citenamefont {Lee},
  \citenamefont {Grest}, \citenamefont {Grosberg},\ and\ \citenamefont
  {Kremer}}]{Halverson_Kremer_JChemPhys_2011}%
  \BibitemOpen
  \bibfield  {author} {\bibinfo {author} {\bibfnamefont {J.~D.}\ \bibnamefont
  {Halverson}}, \bibinfo {author} {\bibfnamefont {W.~B.}\ \bibnamefont {Lee}},
  \bibinfo {author} {\bibfnamefont {G.~S.}\ \bibnamefont {Grest}}, \bibinfo
  {author} {\bibfnamefont {A.~Y.}\ \bibnamefont {Grosberg}}, \ and\ \bibinfo
  {author} {\bibfnamefont {K.}~\bibnamefont {Kremer}},\ }\href@noop {}
  {\bibfield  {journal} {\bibinfo  {journal} {The Journal of chemical physics}\
  }\textbf {\bibinfo {volume} {134}},\ \bibinfo {pages} {204904} (\bibinfo
  {year} {2011}{\natexlab{a}})}\BibitemShut {NoStop}%
\bibitem [{\citenamefont {Halverson}\ \emph
  {et~al.}(2011{\natexlab{b}})\citenamefont {Halverson}, \citenamefont {Lee},
  \citenamefont {Grest}, \citenamefont {Grosberg},\ and\ \citenamefont
  {Kremer}}]{Halverson_Kremer_JChemPhys_2011_2}%
  \BibitemOpen
  \bibfield  {author} {\bibinfo {author} {\bibfnamefont {J.~D.}\ \bibnamefont
  {Halverson}}, \bibinfo {author} {\bibfnamefont {W.~B.}\ \bibnamefont {Lee}},
  \bibinfo {author} {\bibfnamefont {G.~S.}\ \bibnamefont {Grest}}, \bibinfo
  {author} {\bibfnamefont {A.~Y.}\ \bibnamefont {Grosberg}}, \ and\ \bibinfo
  {author} {\bibfnamefont {K.}~\bibnamefont {Kremer}},\ }\href@noop {}
  {\bibfield  {journal} {\bibinfo  {journal} {The Journal of chemical physics}\
  }\textbf {\bibinfo {volume} {134}},\ \bibinfo {pages} {204905} (\bibinfo
  {year} {2011}{\natexlab{b}})}\BibitemShut {NoStop}%
\bibitem [{\citenamefont {Weeks}\ \emph {et~al.}(1971)\citenamefont {Weeks},
  \citenamefont {Chandler},\ and\ \citenamefont
  {Andersen}}]{Weeks_Andersen_JChemPhys_1971}%
  \BibitemOpen
  \bibfield  {author} {\bibinfo {author} {\bibfnamefont {J.~D.}\ \bibnamefont
  {Weeks}}, \bibinfo {author} {\bibfnamefont {D.}~\bibnamefont {Chandler}}, \
  and\ \bibinfo {author} {\bibfnamefont {H.~C.}\ \bibnamefont {Andersen}},\
  }\href {\doibase 10.1063/1.1674820} {\bibfield  {journal} {\bibinfo
  {journal} {The Journal of Chemical Physics}\ }\textbf {\bibinfo {volume}
  {54}},\ \bibinfo {pages} {5237} (\bibinfo {year} {1971})},\ \Eprint
  {http://arxiv.org/abs/https://doi.org/10.1063/1.1674820}
  {https://doi.org/10.1063/1.1674820} \BibitemShut {NoStop}%
\bibitem [{\citenamefont {Doi}\ and\ \citenamefont
  {Edwards}(1988)}]{Doi_Edwards}%
  \BibitemOpen
  \bibfield  {author} {\bibinfo {author} {\bibfnamefont {M.}~\bibnamefont
  {Doi}}\ and\ \bibinfo {author} {\bibfnamefont {S.}~\bibnamefont {Edwards}},\
  }\href@noop {} {\emph {\bibinfo {title} {The theory of polymer dynamics}}}\
  (\bibinfo  {publisher} {Oxford University Press},\ \bibinfo {year}
  {1988})\BibitemShut {NoStop}%
\bibitem [{\citenamefont {Guo}\ \emph {et~al.}(2020)\citenamefont {Guo},
  \citenamefont {Li}, \citenamefont {Wu}, \citenamefont {He},\ and\
  \citenamefont {Zhang}}]{Guo_Zhang_Polymers_2020}%
  \BibitemOpen
  \bibfield  {author} {\bibinfo {author} {\bibfnamefont {F.}~\bibnamefont
  {Guo}}, \bibinfo {author} {\bibfnamefont {K.}~\bibnamefont {Li}}, \bibinfo
  {author} {\bibfnamefont {J.}~\bibnamefont {Wu}}, \bibinfo {author}
  {\bibfnamefont {L.}~\bibnamefont {He}}, \ and\ \bibinfo {author}
  {\bibfnamefont {L.}~\bibnamefont {Zhang}},\ }\href {\doibase
  10.3390/polym12112659} {\bibfield  {journal} {\bibinfo  {journal} {Polymers}\
  }\textbf {\bibinfo {volume} {12}},\ \bibinfo {pages} {2659} (\bibinfo {year}
  {2020})}\BibitemShut {NoStop}%
\bibitem [{\citenamefont {Thompson}\ \emph {et~al.}(2022)\citenamefont
  {Thompson}, \citenamefont {Aktulga}, \citenamefont {Berger}, \citenamefont
  {Bolintineanu}, \citenamefont {Brown}, \citenamefont {Crozier}, \citenamefont
  {in~'t Veld}, \citenamefont {Kohlmeyer}, \citenamefont {Moore}, \citenamefont
  {Nguyen}, \citenamefont {Shan}, \citenamefont {Stevens}, \citenamefont
  {Tranchida}, \citenamefont {Trott},\ and\ \citenamefont {Plimpton}}]{LAMMPS}%
  \BibitemOpen
  \bibfield  {author} {\bibinfo {author} {\bibfnamefont {A.~P.}\ \bibnamefont
  {Thompson}}, \bibinfo {author} {\bibfnamefont {H.~M.}\ \bibnamefont
  {Aktulga}}, \bibinfo {author} {\bibfnamefont {R.}~\bibnamefont {Berger}},
  \bibinfo {author} {\bibfnamefont {D.~S.}\ \bibnamefont {Bolintineanu}},
  \bibinfo {author} {\bibfnamefont {W.~M.}\ \bibnamefont {Brown}}, \bibinfo
  {author} {\bibfnamefont {P.~S.}\ \bibnamefont {Crozier}}, \bibinfo {author}
  {\bibfnamefont {P.~J.}\ \bibnamefont {in~'t Veld}}, \bibinfo {author}
  {\bibfnamefont {A.}~\bibnamefont {Kohlmeyer}}, \bibinfo {author}
  {\bibfnamefont {S.~G.}\ \bibnamefont {Moore}}, \bibinfo {author}
  {\bibfnamefont {T.~D.}\ \bibnamefont {Nguyen}}, \bibinfo {author}
  {\bibfnamefont {R.}~\bibnamefont {Shan}}, \bibinfo {author} {\bibfnamefont
  {M.~J.}\ \bibnamefont {Stevens}}, \bibinfo {author} {\bibfnamefont
  {J.}~\bibnamefont {Tranchida}}, \bibinfo {author} {\bibfnamefont
  {C.}~\bibnamefont {Trott}}, \ and\ \bibinfo {author} {\bibfnamefont {S.~J.}\
  \bibnamefont {Plimpton}},\ }\href {\doibase 10.1016/j.cpc.2021.108171}
  {\bibfield  {journal} {\bibinfo  {journal} {Comp. Phys. Comm.}\ }\textbf
  {\bibinfo {volume} {271}},\ \bibinfo {pages} {108171} (\bibinfo {year}
  {2022})}\BibitemShut {NoStop}%
\bibitem [{si()}]{si}%
  \BibitemOpen
  \href@noop {} {\emph {\bibinfo {title} {See Supplementary Material at [URL to
  be added by publisher] for simulation details and other
  analysis}}}\BibitemShut {NoStop}%
\bibitem [{\citenamefont {Poier}\ \emph
  {et~al.}(2016{\natexlab{a}})\citenamefont {Poier}, \citenamefont {Ba{\u
  c}ov{\' a}}, \citenamefont {Moreno}, \citenamefont {Likos},\ and\
  \citenamefont {Blaak}}]{Poier_Blaak_SoftMatter_2016}%
  \BibitemOpen
  \bibfield  {author} {\bibinfo {author} {\bibfnamefont {P.}~\bibnamefont
  {Poier}}, \bibinfo {author} {\bibfnamefont {P.}~\bibnamefont {Ba{\u c}ov{\'
  a}}}, \bibinfo {author} {\bibfnamefont {A.~J.}\ \bibnamefont {Moreno}},
  \bibinfo {author} {\bibfnamefont {C.~N.}\ \bibnamefont {Likos}}, \ and\
  \bibinfo {author} {\bibfnamefont {R.}~\bibnamefont {Blaak}},\ }\href
  {\doibase 10.1039/C6SM00430J} {\bibfield  {journal} {\bibinfo  {journal}
  {Soft Matter}\ }\textbf {\bibinfo {volume} {12}},\ \bibinfo {pages} {4805}
  (\bibinfo {year} {2016}{\natexlab{a}})}\BibitemShut {NoStop}%
\bibitem [{\citenamefont {Poier}\ \emph
  {et~al.}(2015{\natexlab{a}})\citenamefont {Poier}, \citenamefont {Likos},
  \citenamefont {Moreno},\ and\ \citenamefont
  {Blaak}}]{Poier_Blaak_Macromol_2015}%
  \BibitemOpen
  \bibfield  {author} {\bibinfo {author} {\bibfnamefont {P.}~\bibnamefont
  {Poier}}, \bibinfo {author} {\bibfnamefont {C.~N.}\ \bibnamefont {Likos}},
  \bibinfo {author} {\bibfnamefont {A.~J.}\ \bibnamefont {Moreno}}, \ and\
  \bibinfo {author} {\bibfnamefont {R.}~\bibnamefont {Blaak}},\ }\href
  {\doibase 10.1021/acs.macromol.5b00603} {\bibfield  {journal} {\bibinfo
  {journal} {Macromolecules}\ }\textbf {\bibinfo {volume} {48}},\ \bibinfo
  {pages} {4983} (\bibinfo {year} {2015}{\natexlab{a}})}\BibitemShut {NoStop}%
\bibitem [{\citenamefont {Slimani}\ \emph {et~al.}(2014)\citenamefont
  {Slimani}, \citenamefont {Bacova}, \citenamefont {Bernabei}, \citenamefont
  {Narros}, \citenamefont {Likos},\ and\ \citenamefont
  {Moreno}}]{slimani2014cluster}%
  \BibitemOpen
  \bibfield  {author} {\bibinfo {author} {\bibfnamefont {M.~Z.}\ \bibnamefont
  {Slimani}}, \bibinfo {author} {\bibfnamefont {P.}~\bibnamefont {Bacova}},
  \bibinfo {author} {\bibfnamefont {M.}~\bibnamefont {Bernabei}}, \bibinfo
  {author} {\bibfnamefont {A.}~\bibnamefont {Narros}}, \bibinfo {author}
  {\bibfnamefont {C.~N.}\ \bibnamefont {Likos}}, \ and\ \bibinfo {author}
  {\bibfnamefont {A.~J.}\ \bibnamefont {Moreno}},\ }\href@noop {} {\bibfield
  {journal} {\bibinfo  {journal} {ACS Macro Letters}\ }\textbf {\bibinfo
  {volume} {3}},\ \bibinfo {pages} {611} (\bibinfo {year} {2014})}\BibitemShut
  {NoStop}%
\bibitem [{\citenamefont {Bernabei}\ \emph {et~al.}(2013)\citenamefont
  {Bernabei}, \citenamefont {Bacova}, \citenamefont {Moreno}, \citenamefont
  {Narros},\ and\ \citenamefont {Likos}}]{bernabei2013fluids}%
  \BibitemOpen
  \bibfield  {author} {\bibinfo {author} {\bibfnamefont {M.}~\bibnamefont
  {Bernabei}}, \bibinfo {author} {\bibfnamefont {P.}~\bibnamefont {Bacova}},
  \bibinfo {author} {\bibfnamefont {A.~J.}\ \bibnamefont {Moreno}}, \bibinfo
  {author} {\bibfnamefont {A.}~\bibnamefont {Narros}}, \ and\ \bibinfo {author}
  {\bibfnamefont {C.~N.}\ \bibnamefont {Likos}},\ }\href@noop {} {\bibfield
  {journal} {\bibinfo  {journal} {Soft Matter}\ }\textbf {\bibinfo {volume}
  {9}},\ \bibinfo {pages} {1287} (\bibinfo {year} {2013})}\BibitemShut
  {NoStop}%
\bibitem [{\citenamefont {Poier}\ \emph
  {et~al.}(2015{\natexlab{b}})\citenamefont {Poier}, \citenamefont {Likos},
  \citenamefont {Moreno},\ and\ \citenamefont {Blaak}}]{poier2015anisotropic}%
  \BibitemOpen
  \bibfield  {author} {\bibinfo {author} {\bibfnamefont {P.}~\bibnamefont
  {Poier}}, \bibinfo {author} {\bibfnamefont {C.~N.}\ \bibnamefont {Likos}},
  \bibinfo {author} {\bibfnamefont {A.~J.}\ \bibnamefont {Moreno}}, \ and\
  \bibinfo {author} {\bibfnamefont {R.}~\bibnamefont {Blaak}},\ }\href@noop {}
  {\bibfield  {journal} {\bibinfo  {journal} {Macromolecules}\ }\textbf
  {\bibinfo {volume} {48}},\ \bibinfo {pages} {4983} (\bibinfo {year}
  {2015}{\natexlab{b}})}\BibitemShut {NoStop}%
\bibitem [{\citenamefont {Poier}\ \emph
  {et~al.}(2016{\natexlab{b}})\citenamefont {Poier}, \citenamefont
  {Ba{\v{c}}ov{\'a}}, \citenamefont {Moreno}, \citenamefont {Likos},\ and\
  \citenamefont {Blaak}}]{poier2016anisotropic}%
  \BibitemOpen
  \bibfield  {author} {\bibinfo {author} {\bibfnamefont {P.}~\bibnamefont
  {Poier}}, \bibinfo {author} {\bibfnamefont {P.}~\bibnamefont
  {Ba{\v{c}}ov{\'a}}}, \bibinfo {author} {\bibfnamefont {A.~J.}\ \bibnamefont
  {Moreno}}, \bibinfo {author} {\bibfnamefont {C.~N.}\ \bibnamefont {Likos}}, \
  and\ \bibinfo {author} {\bibfnamefont {R.}~\bibnamefont {Blaak}},\
  }\href@noop {} {\bibfield  {journal} {\bibinfo  {journal} {Soft Matter}\
  }\textbf {\bibinfo {volume} {12}},\ \bibinfo {pages} {4805} (\bibinfo {year}
  {2016}{\natexlab{b}})}\BibitemShut {NoStop}%
\bibitem [{\citenamefont {Schnyder}\ \emph {et~al.}(2015)\citenamefont
  {Schnyder}, \citenamefont {Spanner}, \citenamefont {H{\"o}fling},
  \citenamefont {Franosch},\ and\ \citenamefont
  {Horbach}}]{schnyder2015rounding}%
  \BibitemOpen
  \bibfield  {author} {\bibinfo {author} {\bibfnamefont {S.~K.}\ \bibnamefont
  {Schnyder}}, \bibinfo {author} {\bibfnamefont {M.}~\bibnamefont {Spanner}},
  \bibinfo {author} {\bibfnamefont {F.}~\bibnamefont {H{\"o}fling}}, \bibinfo
  {author} {\bibfnamefont {T.}~\bibnamefont {Franosch}}, \ and\ \bibinfo
  {author} {\bibfnamefont {J.}~\bibnamefont {Horbach}},\ }\href@noop {}
  {\bibfield  {journal} {\bibinfo  {journal} {Soft Matter}\ }\textbf {\bibinfo
  {volume} {11}},\ \bibinfo {pages} {701} (\bibinfo {year} {2015})}\BibitemShut
  {NoStop}%
\bibitem [{\citenamefont {Schnyder}\ and\ \citenamefont
  {Horbach}(2018)}]{schnyder2018crowding}%
  \BibitemOpen
  \bibfield  {author} {\bibinfo {author} {\bibfnamefont {S.~K.}\ \bibnamefont
  {Schnyder}}\ and\ \bibinfo {author} {\bibfnamefont {J.}~\bibnamefont
  {Horbach}},\ }\href@noop {} {\bibfield  {journal} {\bibinfo  {journal}
  {Physical Review Letters}\ }\textbf {\bibinfo {volume} {120}},\ \bibinfo
  {pages} {078001} (\bibinfo {year} {2018})}\BibitemShut {NoStop}%
\bibitem [{\citenamefont {Vaibhav}\ \emph {et~al.}(2022)\citenamefont
  {Vaibhav}, \citenamefont {Horbach},\ and\ \citenamefont
  {Chaudhuri}}]{vaibhav2022finite}%
  \BibitemOpen
  \bibfield  {author} {\bibinfo {author} {\bibfnamefont {V.}~\bibnamefont
  {Vaibhav}}, \bibinfo {author} {\bibfnamefont {J.}~\bibnamefont {Horbach}}, \
  and\ \bibinfo {author} {\bibfnamefont {P.}~\bibnamefont {Chaudhuri}},\
  }\href@noop {} {\bibfield  {journal} {\bibinfo  {journal} {The Journal of
  Chemical Physics}\ }\textbf {\bibinfo {volume} {156}},\ \bibinfo {pages}
  {244501} (\bibinfo {year} {2022})}\BibitemShut {NoStop}%
\bibitem [{Note1()}]{Note1}%
  \BibitemOpen
  \bibinfo {note} {The effective diameter of these tubes ($d_{\protect \text
  {cav}}$) is proportional to the radius of gyration of the B-type rings
  ($R^B_g$), i.e., $d_{\protect \text {cav}} \sim 2\times 14.5\sigma $, which
  is sufficiently large to allow the directed motion of the flexible
  rings}\BibitemShut {NoStop}%
\bibitem [{\citenamefont {Tuinier}\ \emph {et~al.}(2003)\citenamefont
  {Tuinier}, \citenamefont {Rieger},\ and\ \citenamefont
  {De~Kruif}}]{tuinier2003depletion}%
  \BibitemOpen
  \bibfield  {author} {\bibinfo {author} {\bibfnamefont {R.}~\bibnamefont
  {Tuinier}}, \bibinfo {author} {\bibfnamefont {J.}~\bibnamefont {Rieger}}, \
  and\ \bibinfo {author} {\bibfnamefont {C.}~\bibnamefont {De~Kruif}},\
  }\href@noop {} {\bibfield  {journal} {\bibinfo  {journal} {Advances in
  colloid and interface science}\ }\textbf {\bibinfo {volume} {103}},\ \bibinfo
  {pages} {1} (\bibinfo {year} {2003})}\BibitemShut {NoStop}%
\end{thebibliography}
\end{document}